\documentclass[aps,pre,twocolumn,superscriptaddress]{revtex4-1}

\usepackage{palatino}
\usepackage{amsmath}
\usepackage{amssymb}
\usepackage{graphicx}
\usepackage{verbatim}
\usepackage[colorlinks=true,citecolor=blue,linkcolor=blue]{hyperref}

\graphicspath{{figures/}}

\newcommand{\bfv}[1]{{\mbox{\boldmath{$#1$}}}}
\newcommand{\bfm}[1]{{\bf #1}}          

\newcommand{\expect}[1]{\left \langle #1 \right \rangle} 

\newcommand{\x}{\bfv{x}}

\newcommand{\T}{\mathrm{T}}                                

\begin{document}


\title{Reweighting from the mixture distribution as a better way to describe the Multistate Bennett Acceptance Ratio}

 \author{Michael R. Shirts}
  \email{michael.shirts@colorado.edu}
 \affiliation{Department of Chemical and Biological Engineering, University of Colorado Boulder, Boulder, CO 80309}

\begin{abstract}The multistate Bennett Acceptance ratio is provably the
  lowest variance unbiased estimator of both free energies and
  ensemble averages, and has a number of important advantages over
  previous methods, such as WHAM. Despite its advantages, the original
  MBAR paper was rather dense and mathematically complicated, limiting
  the extent to which people could expand and apply it. We present
  here a different way to think about MBAR that is much more intuitive
  and makes it clearer why the method works so well.
\end{abstract}
\date{\today} 

\maketitle


\section{Introduction}

Several years ago, we derived the multistate Bennett Acceptance ratio
(MBAR)~\cite{shirts-chodera:jcp:2008:mbar}, an approach to compute free energies and expectation averages
computed from samples from multiple thermodynamics states.  Based on
results in statistical
inference,~\cite{vardi:1985,gill:1988,kong:2003,tan:2004}, this method
is provably the lowest variance unbiased estimators of both free
energies and ensemble averages.  Additionally, it has a number of
important advantages over previous methods.  Multiple
histogram techniques, such as WHAM, rely on histograms of width
sufficient to contain many samples, which introduces a bias that can be
substantial and often difficult to
assess~\cite{kobrak:j-comput-chem:2003:histogram-error}.
Additionally, using multiple histogram techniques makes it very
difficult to compute error estimates in the free energies or ensemble
averages obtained.  MBAR uses no histograms, estimating quantities
using the samples directly, and thus has no histogram bias.  It is
important to note that equation for free energies itself was derived
by Souaille and Roux in 2001~\cite{souaille:2001a} as the limit of WHAM as
bins go to zero width, so the equation itself was not new, but the
context, connection to statistics, proofs of minimum variance, and
useful formula for uncertainties in the free energies and observables
were novel.

MBAR reduces exactly to Bennett's acceptance ratio formula in the case
of only two states~\cite{bennett:1976}, motivating our original name.
BAR, therefore, does have provably lowest variance for the free energy
calculation between two states, suffers no histogram bias, and has a
robust variance estimate.  However, it cannot be used to directly
compute free energies using samples for multiple states all together
(only in pairs), and cannot be used to compute ensemble averages in
its standard formulation.  MBAR, as we showed in our original paper,
follows the same general mathematical derivation as BAR, yielding
provably lowest variance free energies as well as robust variance
estimates, but extended to multiple states, and the mathematical
framework allows computing the expectation values of observables as
well.

Despite all these advantages, the original MBAR paper was rather dense
and mathematically complicated, somewhat suppressing the understanding
of the approach and therefore, its use by other researchers for
practical problems. We present here a different way to think about
MBAR that is hopefully much more intuitive and makes it clearer why it
works well.

\section{Original mathematical definition of MBAR}

First, some definitions. Suppose we obtain $N_i$ uncorrelated
equilibrium samples~\cite{MLEmeasure_foot2} from each of $K$
thermodynamics states within the same class of ensemble (such as all
NVT, all NPT, all $\bfm{\mu}$VT, etc.).  Each state is characterized
by a specified combination of inverse temperature, potential energy
function, pressure, and/or chemical potential(s), depending upon the
ensemble.  We define the \emph{reduced potential function} $u_i(\x)$
for state $i$ to be
\begin{eqnarray}
u_i(\x) &=& \beta_i [ U_i(\x) + p_i V(\x) + \bfv{\mu}_i^\T \bfv{n}(\x) ] \label{equation:reduced-potential}
\end{eqnarray}
where $\x \in \bfm{\Gamma}$ denotes the configuration of the system
within a configuration space $\bfm{\Gamma}$, with volume $V(\x)$ (in
the case of a constant pressure ensemble) and $\bfv{n}(\x)$ the number
of molecules of each of $M$ components of the system (in the case of a
grand or semigrand ensemble).  For each state $i$, $\beta_i$ is the
inverse temperature, $U_i(\x)$ the potential energy function (which
may include biasing weights), $p_i$ the external pressure, and
$\bfv{\mu}_i$ the vector of chemical potentials of the $M$ system
components.  This formalism allows a very large number of different
situations to be described by the same mathematics.

In all of these situations, configurations $\{\x_{in}\}_{n=1}^{N_i}$
from state $i$ are sampled from the probability distribution
\begin{eqnarray}
p_i(\x) = c_i^{-1} q_i(\x) \: &;& \: c_i = \int_{\bfm{\Gamma}} d\x \, q_i(\x)
\end{eqnarray}
where $q_i(\x)$ is here nonnegative and represents an unnormalized
density function. $c_i$ is the (generally unknown) normalization
constant (which of course, in statistical mechanics is simply the
\emph{partition function}). In samples obtained from standard
Metropolis Monte Carlo or molecular dynamics simulations or from
experiment, this unnormalized density $q_i(\x)$ is simply the
Boltzmann weight $\exp[-u_i(\x)]$ but the general math allows it to be
an arbitrary probability distribution, such as found in simulations
employing non-Boltzmann weights, such as multicanonical
simulations~\cite{mezei:j-comp-phys:1987:muca} or Tsallis
statistics~\cite{tsallis:j-stat-phys:1988:tsallis-statistics}.

What we want is an efficient and simple estimator for the difference
in dimensionless free energies
\begin{eqnarray}
\Delta f_{ij} &\equiv& f_j - f_i = - \ln \frac{c_j}{c_i} = - \ln \frac{\int_\bfm{\Gamma} d\x \, q_j(\x)}{\int_\bfm{\Gamma} d\x \, q_i(\x)} \label{equation:dimensionless-free-energy}
\end{eqnarray}
where the $f_i$ are related to the dimensionalized free energies $F_i$ by $f_i = \beta_i F_i$.  

We would also like to have useful estimators of the equilibrium
expectations of some observable (energy, volume, pair distances, etc.)
of the coordinates $O(\x)$.
\begin{eqnarray}
\expect{O}_i &\equiv& \int_{\bfm{\Gamma}} d\x \, p_i(\x) \, O(\x) = \frac{\int_\bfm{\Gamma} d\x \, O(\x) \, q_i(\x)}{\int_\bfm{\Gamma} d\x \, q_i(\x)} .\label{equation:equilibrium-expectation}
\end{eqnarray}

The original MBAR paper~\cite{shirts-chodera:jcp:2008:mbar} presented
a way to estimate these ratios of normalization constants through the
identity
\begin{eqnarray}
c_i \expect{\alpha_{ij} q_j}_i = \int_{\bfm{\Gamma}} d\x \, q_i(\x) \, \alpha_{ij}(\x) \, q_j(\x)  =  c_j \expect{\alpha_{ij} q_i}_j
\label{equation:Bennett_starting}
\end{eqnarray}
which holds for arbitrary choice of functions $\alpha_{ij}(\x)$,
provided the $c_i$ are nonzero.  It is possible to construct the
$\alpha_{ij}(x)$ which leads to minimum variance in the free energies,
and obtain a self-consistent nonlinear equation for the $c_i$ that has
a unique solution, up to a multiplicative constant:
\begin{eqnarray}
\hat{c}_i = \sum_{n=1}^N \frac{ q_i(\x_n)}{\sum\limits_{k=1}^K N_k \, \hat{c}_k^{-1} \,\ q_k(\x_n)},
\end{eqnarray}
where $N = \sum_{k=1}^K N_k$. One can also construct an estimate of the
asymptotic error of the estimate of $c_i(x)$.~\citep{shirts-chodera:jcp:2008:mbar}

In terms of free energies, this becomes.
\begin{eqnarray}
\hat f_i &=& - \ln \sum_{j=1}^K \sum_{n=1}^{N_j} \frac{\exp[-u_i(\x_{jn})]}{\sum\limits_{k=1}^K N_k \, \exp[\hat{f}_k - u_k(\x_{jn})]} \label{equation:estimator-of-free-energies}
\end{eqnarray}
Again, because the normalization constants are only determined up to a
multiplicative constant, the estimated free energies $\hat{f}_i$ are
determined uniquely only up to an additive constant, so only
differences $\Delta \hat{f}_{ij} = \hat{f}_j - \hat{f}_i$ will be
meaningful.  Estimators of the uncertainties in $\Delta \hat{f}_{ij}$
can again be derived and are given in a previous
paper~\cite{shirts-chodera:jcp:2008:mbar}. 

For a free energy $f_i$ of a state from which no samples are collected,
then we can use the same equation, but we note that the denominator
does not contain that potential, so the free energy does not require
any self-consistent iteration, but instead only requires running
through once, using precomputed denominators.

Once one has computed the normalizing constants (and therefore the free
energies), then one can estimate the equilibrium expectation of any
observable $O(\x)$ that depends only on configuration $\x$ is given by
Eq.~\ref{equation:equilibrium-expectation}. This expectation can be
computed as a ratio of ``normalization'' constants $c_O/c_a$ by
defining the additional function
\begin{eqnarray}
q_O(\x) = O(\x) \, q(\x) 
\end{eqnarray}
While $q_O(\x)$ may no longer be strictly nonnegative, we may still
make use of the equation as long as it does not appear in the
denominator~\cite{doss:2003}, which is the case, since we are not
calculating from this distribution. It's robust and works well, but is
a bit convoluted.  But as we see, there are simpler ways to think
about these expectations (more on that in a bit).

We can write this ratio of integrals as:
\begin{eqnarray}
\langle O_i \rangle = \sum_n O_i(x_n) W_{in} 
\end{eqnarray}
where:
\begin{eqnarray}
W_{ni} = \hat{c}_i^{-1} \, q_i(\x_n) / \sum\limits_{k=1}^K N_k \, \hat{c}_k^{-1} \,\ q_k(\x_n) \label{equation:p_weights} .
\end{eqnarray}
Or, for the Boltzmann distribution:
\begin{eqnarray}
W_{ni} = \frac{\exp[\hat{f}_i-u_i(\x_{n})]}{\sum\limits_{k=1}^K N_k \, \exp[\hat{f}_k - u_k(\x_{n})]}
\end{eqnarray}

Note that we are now indexing the weights by a single index $n =
1,\ldots,N$, rather than a separate index for each state, as,
surprisingly, the association of which samples $\x_n$ came from which
distribution $p_i(\x)$ does not enter into the calculation!  We could
literally forget which state each sample came from, and get the same
answers.  Why is this?

\section{MBAR as importance sampling from the mixture distribution}

A much simpler way to interpret MBAR which provides significant
insight is as reweighting from a mixture distribution. To describe
this well, we first need to review the idea of importance sampling.

One can calculate averages of any observable $O(\vec{x})$ with respect
to a normalized probability distribution $p_i(\vec{x})$ by integrating over the
support of that distribution.
\[ \langle O \rangle_{i} = \int_{\Gamma} O(\vec{x}) p_i(\vec{x}) d\vec{x}\]
Where the subscript $i$ indicates that the average is with respect
to the distribution $p_i(\vec{x})$, and $\Gamma$ is the phase space volume
we integrate over. If we pick samples $\vec{x}$ proportional to their
probability $p_i(\vec{x})$, then we can calculate the same averages by
Monte Carlo integration.
\[ \langle O \rangle_{i} = \frac{1}{N}\sum_{n=1}^{N} O(\vec{x}_n) \]
where $N$ is the number of samples collected.

We can divide and multiply by $p_j(\vec{x})$ to find that:
\[ \langle O \rangle_{i} = \int_{\Gamma} O(\vec{x}) \left(\frac{p_i(\vec{x})}{p_j(\vec{x})}\right) p_j(\vec{x}) d\vec{x}\]

Given a well-behaved $p_j(\vec{x})$, if we can generate samples from
the normalized distribution $p_j(\vec{x})$, then we have:
\[ \langle O \rangle_{i} = \frac{1}{N} \sum_{n=1}^N O(\vec{x}_n) \left(\frac{p_i(\vec{x}_n)}{p_j(\vec{x}_n)}\right) \] 
This equation gives us expectations in state $i$, but with samples
from state $j$. 

Note that to do this, we assumed that $p_j(\vec{x})$ is nonzero in any
finite volume of interest in $p_i(\vec{x})$. In most cases, this will
not be relevant, because if $p_j(\vec{x})=0$, we will not collect
samples from it anyway, so it doesn't matter that the ratio is
undefined, and because with standard pair potentials (Coulomb's law,
Lennard-Jones, etc.), $p_j(\vec{x})=0$ is only true at single points
with zero total volume.  There are some complications with hard
spheres with changing radii, but we will not explore the issues there
at this time.

If $p_j(\vec{x})$ is {\em almost} zero where
$p_i(\vec{x})$ has substantial probability density, then the integrals
will eventually converge, but it will take a very large number of
samples. An example of this latter case is in the insertion of a
Lennard-Jones sphere in a dense fluid, with $p_j(\vec{x})$ the
distribution with a zero potential and $p_i(\vec{x})$ is the
distribution with the Lennard-Jones potential present.  Only with
configurations with other fluid particles at the exact center of the
Lennard-Jones sphere is $p_j(\vec{x})=0$, but it is very nearly zero
for a substantial number of configurations with fluid particles near
the center of the sphere, resulting in an insertion is very
inefficient.  A number of alternative techniques have been developed
(such as soft core approaches and staged insertion) to improve the
convergence of integrals in this situation.

If we know our distributions $p_i(\vec{x})$ and $p_j(\vec{x})$ only up
to unknown constants $c_i$ and $c_j$, so that $p_i(\vec{x}) =
c_i^{-1}q_i(\vec{x})$, then it seems we are stuck; how can we deal
with the unknown ratio $c_i/c_j$? We can introduce a trick; we choose
for the observable $O(\vec{x})=1$, which yields (for samples collected
from the unnormalized distribution $q_i(\vec{x})$):

\begin{eqnarray}
\langle 1 \rangle_{i} &=& \frac{1}{N} \sum_{n=1}^{N} \left(1\right) \frac{c_j}{c_i} \left(\frac{q_i(\vec{x}_n)}{q_j(\vec{x}_n)}\right) \nonumber \\
        1  &=& \frac{1}{N} \sum_{n=1}^{N}\frac{c_j}{c_i} \left(\frac{q_i(\vec{x}_n)}{q_j(\vec{x}_n)}\right) \nonumber \\
       \frac{c_i}{c_j}  &=& \frac{1}{N} \sum_{n=1}^{N} \left(\frac{q_i(\vec{x}_n)}{q_j(\vec{x}_n)}\right)
\end{eqnarray}
since the expectation of 1 is always 1, and 1=1 no matter what state
we are in. In the case of Boltzmann-form distributions, then $c_i =
e^{-f_i}$, the generalized free energy, and $q_i(\vec{x}) =
e^{-u_i(\vec{x})}$, which reduces to
\begin{eqnarray}
f_j-f_i  &=& -\ln \frac{1}{N} \sum_{n=1}^N e^{-(u_i(\vec{x}_n)-u_j(\vec{x}_n))}  \nonumber \\
            &=& -\ln \expect{ e^{-(u_i(\vec{x})-u_j(\vec{x}))}}_{j}
\end{eqnarray}
where $\langle \rangle_j$ indicates ensemble average over the sampled distribution $j$, which is of course the standard one-state reweighting method for free
energy calculations first introduced by Zwanzig.

Now, here's the key step.  Assume we have collected $N_k$ samples from
each of $K$ different distributions $p_k(\vec{x}) = c_k^{-1}
q_k(\vec{x})$. We construct a new probability distribution where we
simply throw all $N = \sum_{i=1}^K N_k$ into the same pot. The
probability of drawing a sample from this mixture of distributions is
going to be simply
\begin{equation}
p_m(\vec{x}) = \frac{1}{N}\sum_{k=1}^K N_k p_k(\vec{x})
\end{equation}
because there is a $\frac{N_k}{N}$ chance of getting a sample from
each of the $K$ distributions. Once we have a sample from that
distribution, the probability distribution is just $p_k(\vec{x})$, as
we have seen.

We call this entire set of samples, together, a {\em mixture
  distribution}, since it involves mixing together samples from all
$K$ distributions.  If each of the individual $p_k(\vec{x})$
distributions is normalized, then it is easy to verify that the
overall mixture distribution $p_m(\vec{x})$ must also normalized.

\subsection*{A visual example}

\begin{figure*}
\begin{tabular}{ccc}
\includegraphics[width=0.33\textwidth]{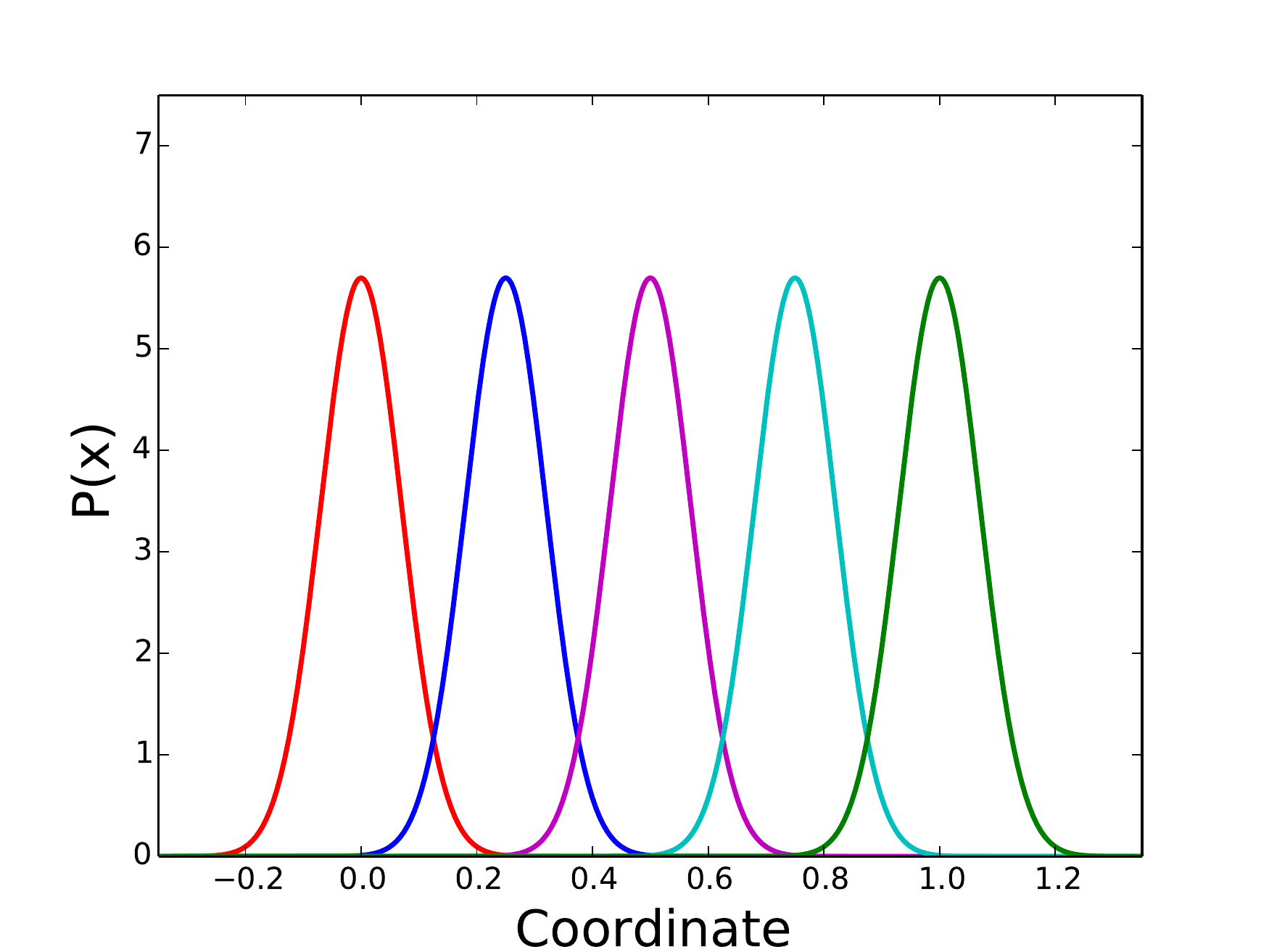}&
\includegraphics[width=0.33\textwidth]{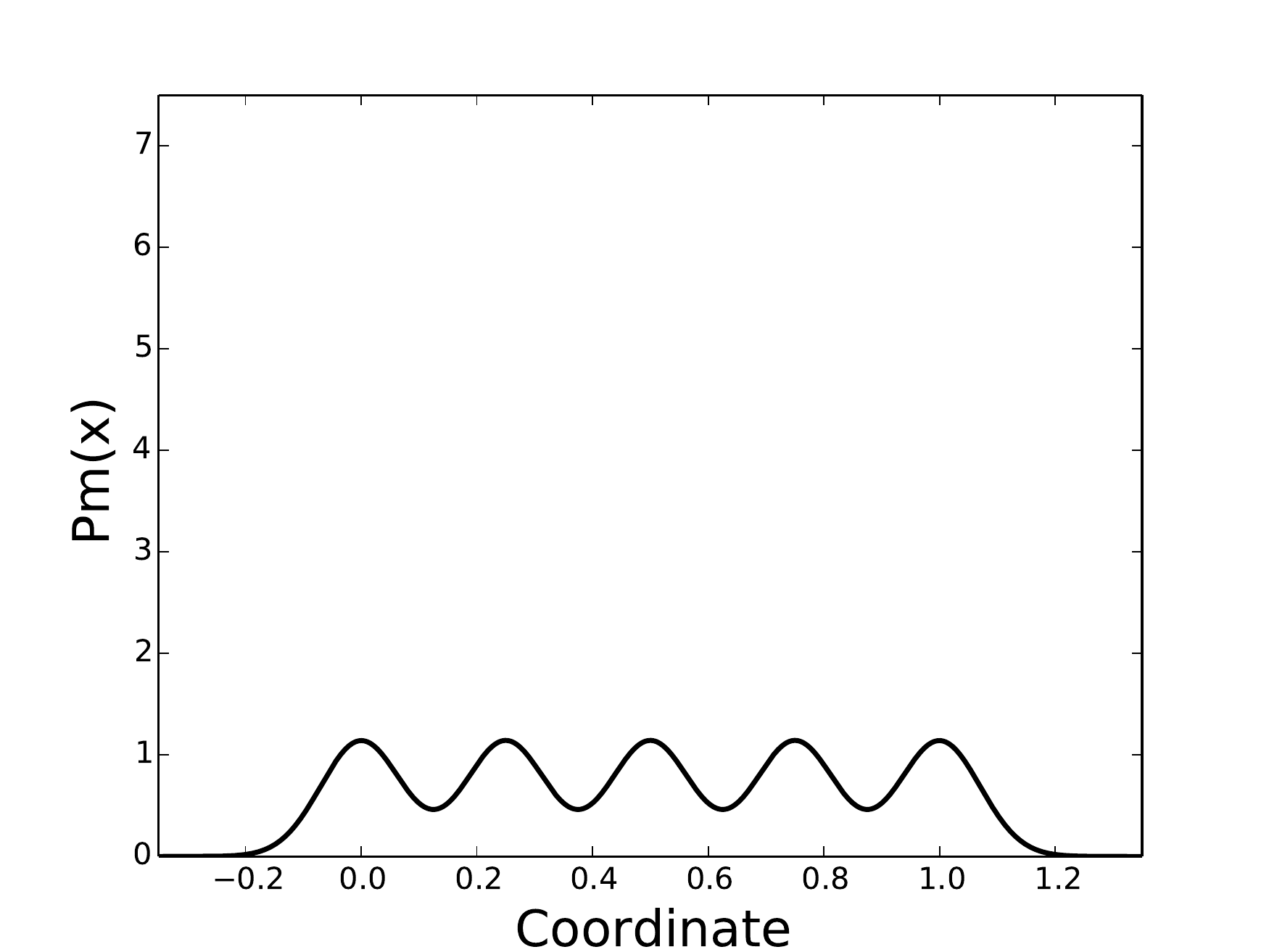}&
\includegraphics[width=0.33\textwidth]{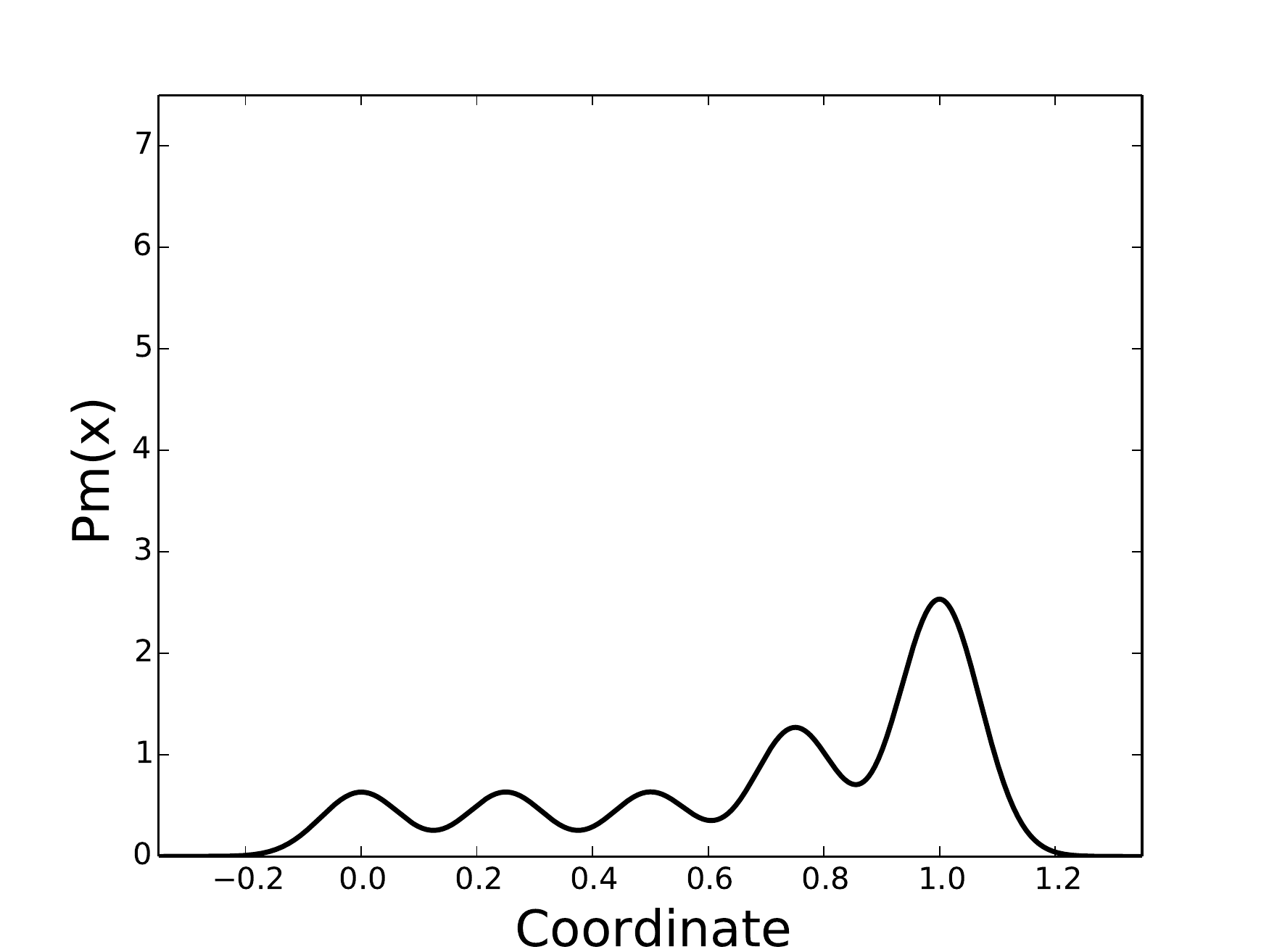} \\
a & b & c
\end{tabular}
\caption{(a) Five normalized probability distributions, that together
  span the entire coordinate range of interest. (b) a mixture
  distribution $p_m(x)$ with even number of samples of each of five
  distributions. (c) a mixture distribution $p_m(x)$ with distribution
  of samples with ratio 1:1:1:2:4. The mixture distributions have
  nonnegligible sampling across the entire configuration range of
  interest, meaning that we can easily reweight to samples at any of
  these configurations. All distributions are normalized.}
\label{figure:mixture}
\end{figure*}

Let's look at a visual example of how one constructs this mixture
distribution (Fig.~\ref{figure:mixture}).  Let's say we have a
one-dimensional degree of freedom, and we collect an equal amount of
data from a series of distributions $p_K$ that span this degree of
freedom.  We gather data from all five distributions
(Fig.~\ref{figure:mixture}a), and pool them together
(Fig.~\ref{figure:mixture}b) into our mixture distribution. If there
are an uneven number of samples from each distribution, we simply
weight each distribution by the number of samples from that
distribution (Fig.~\ref{figure:mixture}c).

How do we calculate the normalizing constants $c_i$ and free energies
$f_i$ so we can reweight observables? Assume again that we only know
our distributions $p_i(\vec{x})$ up to an unknown normalizing constant
$c_i$. The mixture distribution can then be expressed by:
\[ p_m(\vec{x}) = \frac{1}{N}\sum N_k c_k^{-1} q_k(\vec{x}) \]

If each of the individual distributions is normalized (i.e. we solve
for $c_k$), then clearly the mixture distribution $p_m(\vec{x})$ must
be normalized.  If we reweight from the mixture distribution to
distribution $i$, defined by $p_i(\vec{x})$ and use the observable
$O(\vec{x}) = 1$ again, then we have:
\begin{eqnarray} 
  1  &=& \frac{1}{N} \sum_{n=1}^{N}\frac{1}{c_i} \left(\frac{q_i(\vec{x})}{p_m(\vec{x})}\right) \nonumber \\
     &=& \frac{1}{N} \sum_{n=1}^N\frac{c_i^{-1} q_i(\vec{x})}{N^{-1}\sum_k N_k c_k^{-1} q_k(\vec{x_n})} \nonumber \\
     &=& \sum_{n=1}^N \frac{c_i^{-1} q_i(\vec{x}_n)}{\sum_k N_k c_k^{-1} q_k(\vec{x}_n)} \nonumber\\
c_i  &=& \sum_{n=1}^N \frac{q_i(\vec{x}_n)}{\sum_k N_k c_k^{-1} q_k(\vec{x}_n)}
\end{eqnarray}
Where the last equation is simply an algebraic relationship.  There
will be $K$ equations, one for each of the $K$
distributions.  We then have a system of $K$ equations that can be
solved for the $c_k$.  Although there are $K$ equations, there are
only $K-1$ independent equations. This can be seen by noting that the
equations are equivalent if all $c_k$ are multiplied by the same
constant, so we must set one number (often $c_1$, though it could be
any of them) to some value (usually 1). In terms of the Boltzmann
distribution, this becomes.
\begin{eqnarray*}
e^{-f_i}  &=& \sum_{n=1}^N \frac{e^{-u_i(\vec{x}_n)}}{\sum_k N_k e^{f_k - u_k(\vec{x}_n)}}
\end{eqnarray*}
Which is exactly the same as the MBAR equations for $f_i$ previously
published.

We can also write this system of equations as:
\begin{equation*}
1 = \sum_{n=1}^N W_{in}
\end{equation*} 
where: 
\begin{eqnarray*}
W_{in} = \frac{c_i^{-1} q_i(\vec{x}_n)}{\sum_k N_k c_k^{-1} q_k(\vec{x}_n)} 
\end{eqnarray*}
which is true for each $i$, but more simply can be seen as:
\begin{eqnarray}
W_{in} = \frac{1}{N}\frac{p_i(\vec{x}_n)}{p_m(\vec{x}_n)}
\end{eqnarray}
And is simply an expression of the fact that the weights for any of
the ensemble must be normalized.

We can then calculate reweighted expectations of observables at any
state from the mixture distribution, which can be written in a number of ways
\begin{eqnarray}
\langle O \rangle_i  &=& \frac{1}{N}\sum_{i=1}^N O(\vec{x}_n) \frac{p_i(\vec{x}_n)}{p_m(\vec{x}_n)} \nonumber \\
                     &=& \sum_{n=1}^N O(\vec{x}_n) \frac{c_i^{-1} q_i(\vec{x}_n)}{\sum N_k c_k^{-1} q_k(\vec{x}_n)} \nonumber \\ 
                     &=& \sum_{n=1}^N O(\vec{x}_n) \frac{e^{f_i-u_i(\vec{x}_n)}}{\sum N_k e^{f_k - u_k(\vec{x}_n)}} \nonumber \\ 
    &=& \sum_{n=1}^N  O(\vec{x}_n) W_{in} 
\end{eqnarray}
which can be seen as reweighting (or importance sampling) to state $i$
from the mixture distribution $p_m(\vec{x})$ This, again, is exactly
the same equation as the previously published MBAR estimator for
observables.

It then becomes clear why we don't need to know which state each
sample is from. If we are reweighting from the mixture distribution,
we throw out all the information from which state each sample comes
from; the calculation only cares about whether the normalized
probability for each individual sample is known.

The idea of reweighting from a mixture distribution is not original to
us; Geyer appears to be the first to introduced this idea in 1994 in
the case of the mixture of two distributions~\citep{geyer:1994} in a
technical report which did not really appear in published journal or
conference literature; this may explain why the idea never caught
hold, especially outside of the statistics community.

Hopefully, this discussion helps demystify the MBAR equations for free
energy, helps make it clearer where they come from, and what they
mean.

\subsection*{Acknowledgments}

Thanks for Ye Mei (East China Normal University), Tommy Foley (Penn
State), and Yuanyuan Zhou (University of Michigan) for noting typos in
previous versions of this document.


\bibliography{mixture}

\begin{thebibliography}{17}
\expandafter\ifx\csname natexlab\endcsname\relax\def\natexlab#1{#1}\fi
\expandafter\ifx\csname bibnamefont\endcsname\relax
  \def\bibnamefont#1{#1}\fi
\expandafter\ifx\csname bibfnamefont\endcsname\relax
  \def\bibfnamefont#1{#1}\fi
\expandafter\ifx\csname citenamefont\endcsname\relax
  \def\citenamefont#1{#1}\fi
\expandafter\ifx\csname url\endcsname\relax
  \def\url#1{\texttt{#1}}\fi
\expandafter\ifx\csname urlprefix\endcsname\relax\def\urlprefix{URL }\fi
\providecommand{\bibinfo}[2]{#2}
\providecommand{\eprint}[2][]{\url{#2}}

\bibitem[{\citenamefont{Shirts and
  Chodera}(2008)}]{shirts-chodera:jcp:2008:mbar}
\bibinfo{author}{\bibfnamefont{M.~R.} \bibnamefont{Shirts}} \bibnamefont{and}
  \bibinfo{author}{\bibfnamefont{J.~D.} \bibnamefont{Chodera}},
  \bibinfo{journal}{J. Chem. Phys.} \textbf{\bibinfo{volume}{129}},
  \bibinfo{pages}{124105} (\bibinfo{year}{2008}).

\bibitem[{\citenamefont{Vardi}(1985)}]{vardi:1985}
\bibinfo{author}{\bibfnamefont{Y.}~\bibnamefont{Vardi}}, \bibinfo{journal}{Ann.
  Stat.} \textbf{\bibinfo{volume}{13}}, \bibinfo{pages}{178}
  (\bibinfo{year}{1985}).

\bibitem[{\citenamefont{Gill et~al.}(1988)\citenamefont{Gill, Vardi, and
  Wellner}}]{gill:1988}
\bibinfo{author}{\bibfnamefont{R.~D.} \bibnamefont{Gill}},
  \bibinfo{author}{\bibfnamefont{Y.}~\bibnamefont{Vardi}}, \bibnamefont{and}
  \bibinfo{author}{\bibfnamefont{J.~A.} \bibnamefont{Wellner}},
  \bibinfo{journal}{Ann. Stat.} \textbf{\bibinfo{volume}{16}},
  \bibinfo{pages}{1069} (\bibinfo{year}{1988}).

\bibitem[{\citenamefont{Kong et~al.}(2003)\citenamefont{Kong, McCullagh, Meng,
  Nicolae, and Tan}}]{kong:2003}
\bibinfo{author}{\bibfnamefont{A.}~\bibnamefont{Kong}},
  \bibinfo{author}{\bibfnamefont{P.}~\bibnamefont{McCullagh}},
  \bibinfo{author}{\bibfnamefont{X.-L.} \bibnamefont{Meng}},
  \bibinfo{author}{\bibfnamefont{D.}~\bibnamefont{Nicolae}}, \bibnamefont{and}
  \bibinfo{author}{\bibfnamefont{Z.}~\bibnamefont{Tan}}, \bibinfo{journal}{J.
  Royal Stat. Soc. B.} \textbf{\bibinfo{volume}{65}}, \bibinfo{pages}{585}
  (\bibinfo{year}{2003}).

\bibitem[{\citenamefont{Tan}(2004)}]{tan:2004}
\bibinfo{author}{\bibfnamefont{Z.}~\bibnamefont{Tan}}, \bibinfo{journal}{J. Am.
  Stat. Assoc.} \textbf{\bibinfo{volume}{99}}, \bibinfo{pages}{1027}
  (\bibinfo{year}{2004}).

\bibitem[{\citenamefont{Kobrak}(2003)}]{kobrak:j-comput-chem:2003:histogram-error}
\bibinfo{author}{\bibfnamefont{M.~N.} \bibnamefont{Kobrak}},
  \bibinfo{journal}{J. Comput. Chem.} \textbf{\bibinfo{volume}{24}},
  \bibinfo{pages}{1437} (\bibinfo{year}{2003}).

\bibitem[{\citenamefont{Souaille and Roux}(2001)}]{souaille:2001a}
\bibinfo{author}{\bibfnamefont{M.}~\bibnamefont{Souaille}} \bibnamefont{and}
  \bibinfo{author}{\bibfnamefont{B.}~\bibnamefont{Roux}},
  \bibinfo{journal}{Comp. Phys. Commun.} \textbf{\bibinfo{volume}{135}},
  \bibinfo{pages}{40} (\bibinfo{year}{2001}).

\bibitem[{\citenamefont{Bennett}(1976)}]{bennett:1976}
\bibinfo{author}{\bibfnamefont{C.~H.} \bibnamefont{Bennett}},
  \bibinfo{journal}{J. Comp. Phys.} \textbf{\bibinfo{volume}{22}},
  \bibinfo{pages}{245} (\bibinfo{year}{1976}).

\bibitem[{MLE()}]{MLEmeasure_foot2}
\bibinfo{note}{A set of uncorrelated configurations can be obtained from a
  correlated time series, such as is generated by a molecular dynamics or
  Metropolis Monte Carlo simulation, by subsampling the timeseries with an
  interval larger than the statistical inefficiency of the reduced potential
  $u_k$ of the timeseries. The statistical inefficiency can be estimated by
  standard
  procedures~\protect{\cite{swope:1982a,flyvbjerg:1989a,janke:2002a,chodera:jctc:2007}}.}

\bibitem[{\citenamefont{Mezei.}(1987)}]{mezei:j-comp-phys:1987:muca}
\bibinfo{author}{\bibfnamefont{E.}~\bibnamefont{Mezei.}}, \bibinfo{journal}{J.
  Comp. Phys.} \textbf{\bibinfo{volume}{68}}, \bibinfo{pages}{237}
  (\bibinfo{year}{1987}).

\bibitem[{\citenamefont{Tsallis}(1988)}]{tsallis:j-stat-phys:1988:tsallis-statistics}
\bibinfo{author}{\bibfnamefont{C.}~\bibnamefont{Tsallis}}, \bibinfo{journal}{J.
  Stat. Phys.} \textbf{\bibinfo{volume}{52}}, \bibinfo{pages}{479}
  (\bibinfo{year}{1988}).

\bibitem[{dos()}]{doss:2003}
\bibinfo{note}{Honi Doss makes this suggestion in the conference discussion
  of~\protect\cite{kong:2003}}.

\bibitem[{\citenamefont{Geyer}(1994)}]{geyer:1994}
\bibinfo{author}{\bibfnamefont{C.~J.} \bibnamefont{Geyer}},
  \bibinfo{type}{Tech. Rep.} \bibinfo{number}{568},
  \bibinfo{institution}{School of Statistics, University of Minnesota},
  \bibinfo{address}{Minneapolis, Minnesota} (\bibinfo{year}{1994}).

\bibitem[{\citenamefont{Swope et~al.}(1982)\citenamefont{Swope, Andersen,
  Berens, and Wilson}}]{swope:1982a}
\bibinfo{author}{\bibfnamefont{W.~C.} \bibnamefont{Swope}},
  \bibinfo{author}{\bibfnamefont{H.~C.} \bibnamefont{Andersen}},
  \bibinfo{author}{\bibfnamefont{P.~H.} \bibnamefont{Berens}},
  \bibnamefont{and} \bibinfo{author}{\bibfnamefont{K.~R.}
  \bibnamefont{Wilson}}, \bibinfo{journal}{J. Chem. Phys.}
  \textbf{\bibinfo{volume}{76}}, \bibinfo{pages}{637} (\bibinfo{year}{1982}).

\bibitem[{\citenamefont{Flyvbjerg and Petersen}(1989)}]{flyvbjerg:1989a}
\bibinfo{author}{\bibfnamefont{H.}~\bibnamefont{Flyvbjerg}} \bibnamefont{and}
  \bibinfo{author}{\bibfnamefont{H.~G.} \bibnamefont{Petersen}},
  \bibinfo{journal}{J. Chem. Phys.} \textbf{\bibinfo{volume}{91}},
  \bibinfo{pages}{461} (\bibinfo{year}{1989}).

\bibitem[{\citenamefont{Janke}(2002)}]{janke:2002a}
\bibinfo{author}{\bibfnamefont{W.}~\bibnamefont{Janke}}, in
  \emph{\bibinfo{booktitle}{Quantum Simulations of Complex Many-Body Systems:
  From Theory to Algorithms}}, edited by
  \bibinfo{editor}{\bibfnamefont{J.}~\bibnamefont{Grotendorst}},
  \bibinfo{editor}{\bibfnamefont{D.}~\bibnamefont{Marx}}, \bibnamefont{and}
  \bibinfo{editor}{\bibfnamefont{A.}~\bibnamefont{Murmatsu}}
  (\bibinfo{publisher}{John von Neumann Institute for Computing},
  \bibinfo{address}{J{\"{u}}lich, Germany}, \bibinfo{year}{2002}),
  vol.~\bibinfo{volume}{10}, pp. \bibinfo{pages}{423--445}.

\bibitem[{\citenamefont{Chodera et~al.}(2007)\citenamefont{Chodera, Swope,
  Pitera, Seok, and Dill}}]{chodera:jctc:2007}
\bibinfo{author}{\bibfnamefont{J.~D.} \bibnamefont{Chodera}},
  \bibinfo{author}{\bibfnamefont{W.~C.} \bibnamefont{Swope}},
  \bibinfo{author}{\bibfnamefont{J.~W.} \bibnamefont{Pitera}},
  \bibinfo{author}{\bibfnamefont{C.}~\bibnamefont{Seok}}, \bibnamefont{and}
  \bibinfo{author}{\bibfnamefont{K.~A.} \bibnamefont{Dill}},
  \bibinfo{journal}{J. Chem. Theor. Comput.} \textbf{\bibinfo{volume}{3}},
  \bibinfo{pages}{26} (\bibinfo{year}{2007}).

\end{thebibliography}
\end{document}